\documentclass[a4paper, english]{llncs}

\usepackage{graphicx}
\usepackage{latexsym}
\usepackage{amssymb, amsmath}
\usepackage{color}
\usepackage[colorlinks, linkcolor={blue}, citecolor={blue},urlcolor={blue},bookmarks,bookmarksopen,bookmarksdepth=2]{hyperref}
\usepackage[noline,noend,algoruled,linesnumbered,noresetcount]{algorithm2e}
\usepackage[font={small,it}]{caption}
\usepackage{subcaption}
\usepackage{wrapfig}
\usepackage[english]{babel}
\usepackage{here}

\DontPrintSemicolon

\def \nlSty {\NlSty}

\SetFuncSty{sc}
\SetDataSty{em}
\SetCommentSty{em}

\SetKwFor{func}{function}{}{end}
\SetKwFor{upon}{upon}{}{end}
\SetKwFor{mutex}{mutex}{}{end}
\SetKwFor{struct}{struct}{}{end}

\SetKw{Break}{break}

\SetKwFunction{Start}{Start}
\SetKwFunction{Commit}{Commit}
\SetKwFunction{Abort}{Abort}
\SetKwFunction{Conflict}{Conflict}
\SetKwFunction{Schedule}{Schedule}
\SetKwFunction{KernelSTMInit}{KernelSTMInit}

\SetKwFunction{Elect}{Elect}
\SetKwFunction{CurrentThreadIdx}{CurrentThread}
\SetKwFunction{STMSharedStructPtr}{STMSharedStructPtr}
\SetKwFunction{InTx}{InTx}
\SetKwFunction{SetInTx}{SetInTx}
\SetKwFunction{First}{First}
\SetKwFunction{Next}{Next}
\SetKwFunction{ClockTick}{ClockTick}
\SetKwFunction{Suspend}{Suspend}
\SetKwFunction{Yield}{Yield}
\SetKwFunction{CreatesCycle}{CreatesCycle}
\SetKwFunction{MoveToWaitQueue}{MoveToWaitQueue}
\SetKwFunction{OnReadyQueue}{OnReadyQueue}
\SetKwFunction{MoveToReadyQueue}{MoveToReadyQueue}
\SetKwFunction{GetNextFreeSTMStruct}{GetNextFreeSTMStruct}
\SetKwFunction{InitQueue}{InitQueue}
\SetKwFunction{Wait}{Wait}
\SetKwFunction{Release}{Release}
\SetKwFunction{ReleaseAll}{ReleaseAll}
\SetKwFunction{Renice}{Renice}

\SetKwData{cf}{$\mathcal{C}$}
\SetKwData{q}{$\mathcal{Q}$}
\SetKwData{qf}{$\q.first$}
\SetKwData{tr}{$tx.thr$}
\SetKwData{trtx}{$tx.thr.tx$}
\SetKwData{trw}{$tx.thr.wait$}
\SetKwData{tx}{$tx$}
\SetKwData{trp}{$tx'\!.thr$}
\SetKwData{trpw}{$tx'\!.thr.wait$}
\SetKwData{txp}{$tx'$}
\SetKwData{ttx}{$t.tx$}
\SetKwData{N}{$MaxThread$}
\SetKwData{t1}{$t1$}

\SetKwBlock{Block}{}{}

\newcommand{\remove}[1]{}

\newcommand{\lref}[1]{line~\nlSty{\ref{#1}}}

\newcommand{\llref}[2]{lines~\nlSty{\ref{#1}}--\nlSty{\ref{#2}}}

\newcommand{\tab}{\hspace*{2em}}
\begin{document}

\title{Lightweight Contention Management for Efficient Compare-and-Swap
Operations}


\author{Dave Dice\inst{1}, Danny Hendler\inst{2} and Ilya Mirsky\inst{2}}

\institute{Sun Labs at Oracle \and
           Ben-Gurion University of the Negev}

\maketitle

\tolerance=700
\hyphenation{Atomic-Reference}
\hyphenation{imple-mentation imple-mentations}
\hyphenation{conc-urrency}
\hyphenation{invoc-ations}
\hyphenation{manage-ment}
\hyphenation{conten-tion}
\hyphenation{algo-rithm algo-rithms}
\hyphenation{approxi-mately}
\hyphenation{magni-tude}
\hyphenation{specul-ative}

\begin{abstract}

Many concurrent data-structure implementations -- both blocking and non-blocking -- use the well-known \emph{compare-and-swap} (CAS) operation, supported in hardware by most modern multiprocessor architectures for inter-thread synchronization.

\tab A key weakness of the CAS operation is its performance in the presence of memory contention. When multiple threads concurrently attempt to apply CAS operations to the same s\textit{}hared variable, at most a single thread will succeed in changing the shared variable's value and the CAS operations of all other threads will fail. Moreover, significant degradation in performance occurs when variables manipulated by CAS become contention ``hot spots'', since failed CAS operations congest the interconnect and memory devices and slow down successful CAS operations.

\remove{In the context of Lock-free data structures, employing CAS in a \emph{read--CAS--retry\_if\_failed} scheme, failures in CAS lead to excessive computations.} 

\tab In this work we study the following question: \emph{can software-based contention management improve the efficiency of hardware-provided CAS operations?} In other words, can a software contention management layer, encapsulating invocations of hardware CAS instructions, improve the performance of CAS-based concurrent data-structures?

\tab To address this question, we conduct what is, to the best of our knowledge, the first study on the impact of contention management algorithms on the efficiency of the CAS operation.

\tab We implemented several Java classes, which extend Java's \emph{AtomicReference} class, that encapsulate calls to native CAS with simple contention management mechanisms tuned for different hardwares. A key property of our algorithms is the support for an almost-transparent interchange with Java's \emph{AtomicReference} objects, used in implementat\-ions of concurrent data structures. We then evaluate the impact of these algorithms on both a synthetic micro-benchmark and on CAS-based concurrent implementations of widely-used data-structures such as stacks and queues.

\tab Our performance evaluation establishes that lightweight contention management support can greatly improve performance under medium and high contention levels while typically incurring only small overhead when contention is low. In some cases, applying efficient contention management for CAS operations used by a simpler data-structure implementation yields better performance than highly optimized implementations of that data-structure that use native CAS operations directly.

\remove{ Many concurrent data-structure implementations -- both blocking and non-blocking -- use the \emph{compare-and-swap} (CAS) instruction, supported in hardware by most modern multiprocessor architectures. CAS is used by numerous concurrent algorithms, since its atomic semantics allows threads to read a shared variable, compute a new value which is a function of the value read, and write the new value back only if the shared variable was not changed in the interim by other, concurrent, threads. As proven in Herlihy's seminal paper \cite{waitfree}, CAS can implement, together with reads and writes, any object in a wait-free manner. }


\end{abstract}

\pagestyle{plain}


\section{Introduction}

Many key problems in shared-memory multiprocessors revolve around the
coordination of access to shared resources and can be captured as
\emph{concurrent data structures} \cite{AW98,Herlihy:2008:AMP:1734069}: abstract
data structures that are concurrently accessed by asynchronous threads.
Efficient concurrent data structure algorithms are key to the scalability of
applications on multiprocessor machines. Devising efficient and scalable
concurrent algorithms for widely-used data structures such as counters (e.g.,
\cite{DBLP:journals/tocs/HerlihyLS95,DBLP:journals/dc/HerlihySW96}), queues
(e.g.,\cite{DBLP:conf/wdag/AfekHM11,DBLP:conf/ppopp/FatourouK12,DBLP:conf/opodis/GidenstamST10,FlatCombining,DBLP:conf/wdag/HendlerIST10,DBLP:journals/cacm/SchererLS09,DBLP:journals/dc/Ladan-MozesS08,DBLP:conf/podc/MichaelS96,DBLP:conf/spaa/MoirNSS05},
stacks (e.g.,\cite{DBLP:conf/ppopp/FatourouK12,FlatCombining,DBLP:journals/jpdc/HendlerSY10}), pools (e.g.,\cite{DBLP:conf/europar/AfekKNS10,DBLP:conf/wdag/BasinFKKP11,DBLP:conf/spaa/GidronKPP12}) and hash tables (e.g.,
\cite{DBLP:conf/sc/GoodmanLJ11,DBLP:conf/wdag/HerlihyST08,DBLP:journals/jacm/ShalevS06,DBLP:journals/sigops/TriplettMW10}),
to name a few, is the focus of intense research.

Modern multiprocessors provide hardware support of atomic read-modify-write
operations in order to facilitate inter-thread coordination and synchroni\-zation.
The \emph{compare-and-swap} (CAS) operation has become the synchronization
primitive of choice for implementing concurrent data structures - both
lock-based and nonblocking \cite{waitfree} - and is supported by hardware in
most contemporary multiprocessor
architectures~\cite{Itanium2001,Sparc,Motorola86}. The CAS operation takes three
arguments: a memory address\footnote{In managed programming languages such as Java,
the memory address is encapsulated by the object on which the CAS operation is
invoked and is therefore not explicitly passed to the CAS operation.}, an old
value, and a new value. If the address stores the old value, it is replaced with
the new value; otherwise it is unchanged. The success or failure of the
operation is then reported back to the calling thread.
CAS is widely available and used since its atomic semantics allow threads to
read a shared variable, compute a new value which is a function of the value
read, and write the new value back only if the shared variable was not changed
in the interim by other, concurrent, threads. As proven in Herlihy's seminal
paper \cite{waitfree}, CAS can implement, together with reads and writes, any
object in a wait-free manner.

A key weakness of the CAS operation, known to both researchers and practitioners
of concurrent programming, is its performance in the presence of memory
contention. When multiple threads concurrently attempt to apply CAS operations
to the same shared variable, typically at most a single thread will succeed in
changing the shared variable's value and the CAS operations of all other threads
will fail. Moreover, significant degradation in performance occurs when
variables manipulated by CAS become contention ``hot spots'', since failed CAS
operations congest the interconnect and memory devices and slow down successful
CAS operations.

\begin{wrapfigure}{r}{0.4\textwidth} \centering
	\vspace{-35pt}
	\includegraphics[scale=0.6]{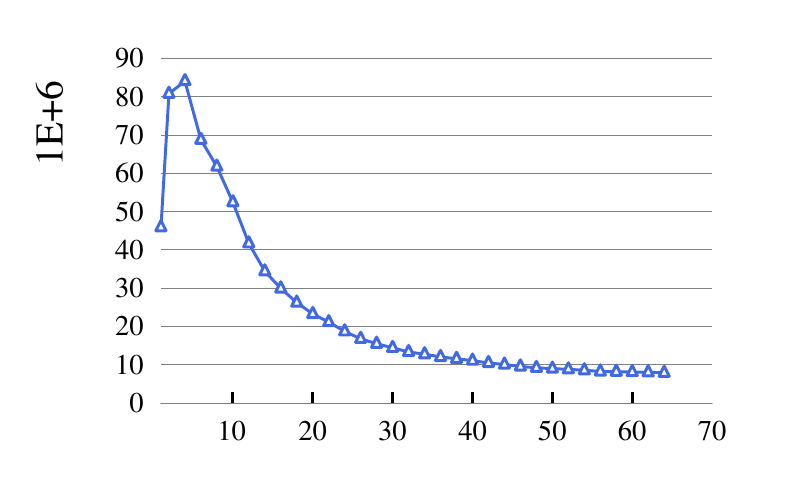}
	\vspace{-25pt}
	\caption{SPARC: Java's CAS}
	\label{figure:SparcNativeCAS}
	\vspace{-30pt}
\end{wrapfigure}

To illustrate this weakness of the CAS operation, Figure
\ref{figure:SparcNativeCAS} shows the results of a simple test, conducted on an
UltraSPARC T2+ (Niagara II) chip, comprising 8 cores,
each multiplexing 8 hardware threads, in which a varying
number of Java threads run for 5 seconds, repeatedly reading the same variable
and then applying CAS operations attempting to change its value.\footnote{We
provide more details on this test in Section \ref{sec:evaluation}.} The number
of successful CAS operations scales from 1 to 4 threads but then quickly
deteriorates, eventually falling to about 16\% of the single thread performance,
less than 9\% of the performance of 4 threads. As we show in Section
\ref{sec:evaluation}, similar performance degradation occurs on Intel's
Xeon and i7 platforms.


In this work we study the following question: \emph{can software-based
contention management improve the efficiency of hardware-provided CAS
operations?} In other words, can a software contention management layer,
encapsulating invocations of hardware CAS instructions, significantly
improve the performance of CAS-based concurrent data-structures?

To address this question, we conduct what is, to the best of our knowledge, the
first study on the impact of contention management algorithms on the efficiency
of the CAS operation. We implemented several Java classes that extend Java's
AtomicReference class, and encapsulate calls to direct CAS by \emph{contention
management layer}. This design allows for an almost transparent plugging of our
classes into existing data structures which make use of Java's
AtomicReference. We then evaluated the impact of these algorithms on the
Xeon, SPARC and i7 platforms by using both a synthetic micro-benchmark and CAS-based concurrent
data-structure implementations of stacks and queues.

We note that the lock-freedom and wait-freedom progress properties
aren't affected by our contention management algorithms since in all of them a thread only waits for a bounded period of time.

The idea of employing contention management and backoff techniques to
improve performance was widely studied in the context of \emph{software transactional memory}
(see, e.g., \cite{Herlihy:1993:TMA:165123.165164,Scherer:2005:ACM:1073814.1073861}) and lock implementations (see, e.g., \cite{Boyd-Wickizer-Locks,mcs}).
Backoff techniques are also used at the higher abstraction level of specific data structure implementations \cite{DBLP:journals/jpdc/HendlerSY10,Herlihy:1993:MIH:161468.161469,Michael:1996:SFP:248052.248106}. However, this approach adds complexity to the design of the data-structure and requires careful per-data structure tuning.
Our approach, of adding contention management (and, specifically, backoff) mechanisms at the CAS instruction level, provides a simple and generic solution, in which tuning can be done \emph{per architecture} rather than per implementation.

Our performance evaluation establishes that lightweight contention manage-ment
support can significantly improve the performance of concurrent data-structure
implementations as compared with direct use of Java's AtomicReference class. Our
CAS contention management algorithms improve the throughput of the concurrent
data-structure implementations we experimented with by a factor of up to 12 for
medium and high contention levels, typically incurring only small overhead in
low contention levels.

We also compared relatively simple data-structure implementations that use our
CAS contention management classes with more complex implementations that employ
data-structure specific optimizations. We have found that, in some cases,
applying efficient contention management at the level of CAS operations, used by simpler and
non-optimized data-structure implementations, yields better performance than
that of highly optimized implementations of the same data-structure that uses
Java's AtomicReference objects directly.

\emph{Our results imply that encapsulating invocations of CAS by lightweight
contention management algorithms is a simple and generic way of significantly
improving the performance of concurrent objects.}

The rest of this paper is organized as follows. We describe the contention
management algorithms we implemented in Section \ref{sec:algorithms}. We report
on our experimental evaluation in Section \ref{sec:evaluation}. We conclude the
paper in Section \ref{sec:discussion} with a short discussion of our results. 
\def\naive{na\"{\i}ve}
\SetKwBlock{Loop}{loop}{end loop}
\SetKwBlock{Proc}{}{}
\SetKw{NCS}{Non Critical Section}
\SetKw{CS}{Critical Section}
\SetKwData{Int}{\textbf{int}}
\SetKwData{Item}{\textbf{item}}
\SetKwData{Void}{\textbf{void}}
\SetKwData{Bool}{\textbf{boolean}}
\SetKwData{multiOp}{\textbf{multiOp}}
\SetKwData{Stack}{\textbf{Stack}}
\SetKwData{Cell}{\textbf{Cell}}
\SetKwData{List}{\textbf{list of}}
\SetKwData{Array}{\textbf{array of}}
\SetKwRepeat{DoWhile}{do}{while}
\SetKwData{Data}{\textbf{Data}}
\SetKw{Of}{of}
\SetKw{Lock}{CAS lock}
\SetKw{Max}{max}
\SetKw{Define}{define}
\SetKw{Init}{initially}
\SetKw{MyFor}{for}
\SetKw{Goto}{goto}
\SetKw{Break}{break}
\SetKw{Await}{await}
\SetKw{True}{true}
\SetKw{False}{false}
\SetKw{Null}{null}
\SetKw{Cont}{continue loop}
\SetKw{Local}{local:}
\SetKw{Shared}{shared}
\SetKw{Global}{global}
\SetKw{Constant}{constant}
\SetKw{AndCond}{$\land$}
\SetKw{OrCond}{$\lor$}

\SetKw{Int}{\textbf{int}}
\SetKw{Bool}{\textbf{boolean}}
\SetKw{Long}{\textbf{long}}
\SetKw{New}{\textbf{new}}
\SetKw{Super}{\textbf{super}}
\SetKw{Volatile}{\textbf{volatile}}
\SetKw{Class}{\textbf{class}}
\SetKw{Private}{\textbf{private}}
\SetKw{Public}{\textbf{public}}
\SetKw{Extends}{\textbf{extends}}

\newcommand{\NIL}{\ensuremath{\bot}}

\section{Contention Management Algorithms}
\label{sec:algorithms}

In this section, we describe the Java CAS contention management algorithms that we implemented and evaluated. These algorithms are implemented as classes that extend the AtomicReference class of the java.util.concurrent.atomic package. Each instance of these classes operates on a specific location in memory and implements the \emph{read} and \emph{CAS} methods.\footnote{None of the methods of AtomicReference are overridden.}

In some of our algorithms, threads need to access per-thread state associated with the object. For example, a thread may keep record of the number of CAS failures it incurred on the object in the past in order to determine how to proceed if it fails again. Such information is stored as an array of per-thread structures. To access this information, threads call a \emph{registerThread} method on the object to obtain an index of an array entry. This thread index is referred to as \texttt{TInd} in the pseudo-code. After registering, a thread may call a \emph{deregisterThread} method on the object to indicate that it is no longer interested in accessing this object and that its entry in this object array may be allocated to another thread.\footnote{An alternative design is to have a global registration/deregistration mechanism so that the TInd index may be used by a thread for accessing several CAS contention-management objects.}

Technically, a thread's TInd index is stored as a thread local variable, using the services of Java's ThreadLocal class. The TInd index may be retrieved within the CAS contention management method implementation. However, in some cases it might be more efficient to retrieve this index at a higher level (for instance, when CAS is called in a loop until it is successful) and to pass it as an argument to the methods of the CAS contention management object.

\subsection{The \texttt{ConstantBackoffCAS} Algorithm}

Algorithm \ref{alg:ConstantBackoff} presents the \texttt{ConstantBackoffCAS} class, which employs the simplest contention management algorithm that we implemented. No per-thread state is required for this algorithm. The \texttt{read} operation simply delegates to the \texttt{get} method of the AtomicReference object to return the current value of the reference (\lref{CONST:callGet}). The \texttt{CAS} operation invokes the \texttt{compareAndSet} method on the AtomicReference superclass, passing to it the \emph{old} and \emph{new} operands (\lref{CONST:invokeCAS}). The \texttt{CAS} operation returns \emph{true} in \lref{CONST:retTrue} if the native CAS succeeded. If the native CAS failed, then the thread busy-waits for a platform-dependent period of time, after which the \texttt{CAS} operation returns (\llref{CONST:wait}{CONST:retFalse}).



\newcommand{\codeConstBackoff}{
\begin{algorithm}[t!]
\scriptsize
\caption{\texttt{ConstBackoffCAS}}
\label{alg:ConstantBackoff}

\Public \Class \textbf{ConstBackoffCAS$<$V$>$} \Extends AtomicReference$<$V$>$\;

\BlankLine

%

\Public V \textbf{read}() {
	\{ \Return get() \nllabel{CONST:callGet} \}
}

\BlankLine

\Public \Bool \textbf{CAS}(V \emph{old}, V \emph{new}) \Proc{
	\uIf {$\lnot$compareAndSet(\emph{old},\emph{new}) \nllabel{CONST:invokeCAS}}
	{
		wait(WAITING\_TIME) \nllabel{CONST:wait}\;
		\Return \False \nllabel{CONST:retFalse}\;
	}
	\lElse
	{
		\Return \True \nllabel{CONST:retTrue}\;
	}
}

\end{algorithm}
}

\codeConstBackoff

\subsection{The \texttt{TimeSliceCAS} Algorithm}

Algorithm \ref{alg:TimeSlice} presents the \texttt{TimeSliceCAS} class, which implements a time-division contention-management algorithm that, under high contention, assigns different time-slices to different threads. Each instance of the class has access to a field $regN$ which stores the number of threads that are currently registered at the object.

The \texttt{read} operation simply delegates to the \texttt{get} method of the AtomicReference class (\lref{TIME:callGet}). The \texttt{CAS} operation invokes the \texttt{compareAndSet} method on the AtomicReference superclass (\lref{TIME:invokeCAS}). If the CAS is successful, the method returns \emph{true} (\lref{TIME:returnTrue}).

If the CAS fails and the number of registered threads exceeds a platform-dependent level CONC (\lref{TIME:shouldLimitConc}), then the algorithm attempts to limit the level of concurrency (that is, the number of threads concurrently attempting CAS on the object) at any given time to at most CONC. This is done as follows. The thread picks a random integer slice number in $\{1, \ldots, \lceil regN/CONC \rceil\}$ (\lref{TIME:Rand}). The length of each time-slice is set to $2^{SLICE}$ nanoseconds, where \emph{SLICE} is a platform-dependent integer. The thread waits until its next time-slice starts and then returns false (\llref{TIME:wait}{TIME:returnFalse}).


\newcommand{\codeTimeSlice}{
\begin{algorithm}[H]
\scriptsize
\caption{\label{alg:TimeSlice}\texttt{TimeSliceCAS}}

\Public \Class \textbf{TimeSliceCAS$<$V$>$} \Extends AtomicReference$<$V$>$\;

\BlankLine

\Public V \textbf{read}() {
	\{ \Return get() \nllabel{TIME:callGet} \}
}

\BlankLine

\Public \Bool \textbf{CAS}(V \emph{old}, V \emph{new}) \Proc{
	\If {\emph{compareAndSet(old,new)} \nllabel{TIME:invokeCAS}}
	{
		\Return \True \nllabel{TIME:returnTrue}
	}
	
	\If {\emph{regN $ > $ CONC} \nllabel{TIME:shouldLimitConc}}
	{
		\Int sliceNum = Random.nextInt($\lceil$regN/CONC$\rceil$) \nllabel{TIME:Rand}\;
		\Repeat{\emph{sliceNum} $ = $ \emph{currentSlice}}
		{
			\nllabel{TIME:wait}
			currentSlice = (System.nanoTime() $ >> $ SLICE) \% $\lceil$regN/CONC$\rceil$\;
		}
	}

 	\Return \False \nllabel{TIME:returnFalse}\;

}

\end{algorithm}
}

\subsection{The \texttt{ExpBackoffCAS} Algorithm}

Algorithm \ref{alg:ExpBackoff} presents the \texttt{ExpBackoffCAS} class, which implements an exponential backoff contention management algorithm. Each instance of this class has a \emph{failures} array, each entry of which -- initialized to 0 -- stores simple per-registered thread statistics about the history of successes and failures of past CAS operations to this object (\lref{EXP:failures}). The \texttt{read} operation simply delegates to the \texttt{get} method of the AtomicReference class (\lref{EXP:callGet}).

The \texttt{CAS} operation invokes the \texttt{compareAndSet} method on the AtomicReference superclass (\lref{EXP:invokeCAS}). If the CAS is successful, then the \texttt{CAS} operation returns \emph{true} (\lref{EXP:returnTrue}).

If the CAS fails, then the thread's entry in the \emph{failures} array is incremented and if its value $f$ is larger than a platform-dependent threshold, the thread waits for a period of time proportional to $2^{min(c \cdot f, m)}$ where $c$ and $m$ are platform-dependent integer algorithm parameters (\llref{EXP:incFailures}{EXP:wait}).


\newcommand{\codeExpBackoff}{
\begin{algorithm}[H]
\scriptsize
\caption{\label{alg:ExpBackoff}\texttt{ExpBackoffCAS}}

\Public \Class \textbf{ExpBackoffCAS$<$V$>$} \Extends AtomicReference$<$V$>$\;
\Private int[] \emph{failures} = \New \Int[MAX\_THREADS] \nllabel{EXP:failures}\;

\BlankLine

%

\Public V \textbf{read}() {
	\{ \Return get() \nllabel{EXP:callGet} \}
}

\BlankLine

\Public \Bool \textbf{CAS}(V \emph{old}, V \emph{new}) \Proc{
	\eIf {compareAndSet(old,new) \nllabel{EXP:invokeCAS}}
	{
	   	\If{failures\emph{[}TInd\emph{]} $>$ 0 \nllabel{EXP:positiveFailures}}
		{
	   		\emph{failures}[\emph{TInd}]$--$\;
	   	}
	   	\BlankLine
		\Return \True \nllabel{EXP:returnTrue}
	}
	{
		\Int \emph{f} = \emph{failures}[\emph{TInd}]$++$ \nllabel{EXP:incFailures}\;
		\If{\emph{f} $>$ \emph{EXP\_THRESHOLD} \nllabel{EXP:wait}}
	   	{
	   		wait($2^{min(c \cdot f, m)}$)\;
	   	}
	   	\BlankLine
	   	\Return \False\;
	}
}

\end{algorithm}
}


\begin{figure}[tb]
\begin{minipage}{0.49\textwidth}
	\codeTimeSlice
\end{minipage}
\begin{minipage}{0.5\textwidth}
	\vspace{-1.8pt}
	\codeExpBackoff
\end{minipage}
\end{figure}

\subsection{The \texttt{MCS-CAS} Algorithm}

With the MCS-CAS algorithm, threads may apply their operations in either \emph{low-contention mode} or \emph{high-contention mode}. Initially, a thread starts operating in low-contention mode, in which it essentially delegates read and CAS operations to the respective methods of the AtomicReference class. When a thread incurs CONTENTION\_THRESHOLD (a platform-dependent constant) consecutive CAS failures on a specific memory location, it reverts to operating in high-contention mode \emph{when accessing this location}.

In high-contention mode, threads that apply CAS operations to the same memory location attempt to serialize their operations by forming a queue determining the order in which their read and CAS operations-pairs will be performed. Threads wait for a bounded period of time within their read operation and proceed to perform the read (and later on the CAS) once the thread that precedes them in the queue (if any) completes its CAS operation.

MCS-CAS implements a variation of the Mellor-Crummey and Scott (MCS) lock algorithm \cite{mcs}. Since we would like to maintain the nonblocking semantics of the CAS operation, a thread $t$ awaits its queue predecessor (if any) for at most a platform-dependent period of time. If this waiting time expires, $t$ proceeds with the read operation without further waiting. If all threads operate in high-contention mode w.r.t. memory location $m$ (and assuming the waiting-time is sufficiently long), then all CAS operations to $m$ will succeed, since each thread may read $m$ and later apply its CAS to $m$ without interruption. In practice, however, threads may apply operations to $m$ concurrently in both low- and high-contention modes and failures may result. After successfully performing a platform-dependent number of CAS operations in high-contention mode, a thread reverts to operating in low-contention mode.

If a thread needs to apply a read that is not followed by a CAS, then it may directly apply the \texttt{get} method of the AtomicReference super-class as this method is not overridden by the MCS-CAS class. There may be situations, however, in which it is not known in advance whether a read will be followed by a CAS and this depends on the value returned by the read. Such scenarios will not compromise the correctness and non-blocking progress of MCS-CAS, but may have adverse effect on performance. This comment applies also to the ArrayBased algorithm described in Section \ref{section:ArrayBased}. The full pseudo-code of the MCS algorithm and its description is provided in appendix A.

\subsection{The \texttt{ArrayBasedCAS} Algorithm}
\label{section:ArrayBased}

The ArrayBased algorithm uses an array-based signalling mechanism, in which a \emph{lock owner} searches for the next entry in the array on which a thread is waiting for permission to proceed with its load-CAS operations in order to signal it. Also in this algorithm, waiting-times are bounded.

There are two key differences between how MCS-CAS and ArrayBasedCAS attempt to serialize read and CAS operations-pairs to a memory location under high contention. First, whereas in MCS-CAS a thread signals its successor after completing a \emph{single} read/CAS operations-pair, with array based a thread performs a multiple, platform-dependent, number of such operations-pairs before signaling other waiting threads.

A second difference is that whereas MCS-CAS forms a dynamic queue in which a thread signals its successor, with array based a thread $t$ that completes its CAS scans the threads records array starting from $t$'s entry for finding a waiting thread to be signaled. This implies that every waiting thread will eventually receive the opportunity to attempt its read/CAS operations-pair.

Since array based does not use a dynamic waiting queue, threads may enter waiting mode and be signaled without having to perform a successful CAS on any of the ArrayBasedCAS data-structures. This is in contrast to MCS-CAS, where a thread must apply a successful CAS to the \emph{tail} variable for joining the waiting queue.

Similarly to MCS-CAS, a thread $t$ waits to be signaled for at most a platform-dependent period of time. If this waiting time expires, $t$ proceeds with its read operation without further waiting. This ensures that array based is nonblocking.

The full pseudo-code of the array-based algorithm and its description appears in Appendix B.

\section{Evaluation}
\label{sec:evaluation}

We conducted our performance evaluation on the SPARC and on Intel's Xeon and i7 multi-core CPUs. The SPARC machine comprises an UltraSPARC T2+ (Niagara II) chip containing 8 cores, each core multiplexing 8 hardware threads, for a total of 64 hardware threads. It runs the 64-bit Solaris 10 operating system with Java SE 1.6.0 update 23. The Xeon machine comprises a Xeon E7-4870 chip, containing 10 cores and hyper-threaded to 20 hardware threads. The i7 machine has an i7-920 CPU comprising 4 cores each supporting 2 hardware threads, for a total of 8 hardware threads. Both Intel machines run the 64-bit Linux 3.2.1 kernel with Java SE 1.6.0 update 25. All tests were conducted with HotSpot in 64-bit \emph{server} mode.

Initially we evaluated our CAS contention management algorithms using a synthetic micro-benchmark and used the results to optimize the platform-dependent parameters used by the algorithms. We then evaluated the impact of our algorithms on implementations of widely-used data structures such as queues and stacks. No explicit threads placement was used.

\remove{ In none of the benchmarks an explicit threads placement was used, but rather it was left to the operating system to handle. It is interesting to notice that the threads placement policy on the SPARC is to first scatter the threads between the cores, and only when the number of running threads exceeds the number of available cores, begin placing multiple threads on the same core.} 

\subsection{The CAS micro-benchmark}

To tune and compare our CAS contention management algorithms, we used the following synthetic \emph{CAS benchmark}. For every concurrency level $k$, varying from 1 to the maximum number of supported hardware threads, $k$ threads repeatedly read the same atomic reference and attempt to CAS its value, for a period of 5 seconds. Before the test begins, each thread generates an array of 128 objects and during the test it attempts to CAS the value of the shared object to a reference to one of these objects, in a round-robin manner. In the course of the test, each thread counts the number of successful CAS operations and these local counters are summed up at the end of the test.

Using the CAS benchmark, we've tuned the parameters used by the algorithms described in Section \ref{sec:algorithms}. The values that were chosen as optimal were those that produced the highest average throughput of all concurrency levels. These values appear in Table \ref{table:summary-tuned-parameters}.\footnote{The values of the WAITING\_TIME and MAX\_WAIT parameters are expressed in milliseconds. Waiting is done by performing a corresponding number of loop iterations.} Figures \ref{figure:XeonJavaCAS}-\ref{figure:SPARCJavaFailures} show the results of the CAS synthetic benchmarks on the three platforms on which we conducted our tests using these optimal parameter values. Each data point is the average of 10 independent executions.

%
%
%
%

\begin{table}[bt]
\tiny
\caption{Summary of tuned algorithm parameters.}
\label{table:summary-tuned-parameters}
\renewcommand{\arraystretch}{1.5}
\vspace{10pt}
\begin{tabular}{|l|l|l|l|}
\hline
		&  			Xeon 			& 			i7		 		&		 Sparc		 		\\
\hline
\hline
CB-CAS	& 	WAITING\_TIME=0.13ms	&	WAITING\_TIME=0.8ms		&	WAITING\_TIME=0.2ms		\\
\hline
		&	EXP\_THRESHOLD=2 		&	EXP\_THRESHOLD=2 		&	EXP\_THRESHOLD=1 		\\
EXP-CAS	&	c = 8              		& 	c = 9              		&	c = 1              		\\
		& 	m = 24             		& 	m = 27             		& 	m = 15             		\\
\hline
		& CONTENTION\_THRESHOLD=8 	& CONTENTION\_THRESHOLD=8 	& CONTENTION\_THRESHOLD=14 	\\
MCS-CAS	& NUM\_OPS = 10,000			& NUM\_OPS = 10,000			& NUM\_OPS = 10				\\
		& MAX\_WAIT = 0.9ms			& MAX\_WAIT = 7.5ms			& MAX\_WAIT = 1ms			\\
\hline
		& CONTENTION\_THRESHOLD=2 	& CONTENTION\_THRESHOLD=2 	& CONTENTION\_THRESHOLD=14 	\\
AB-CAS	& NUM\_OPS = 10,000			& NUM\_OPS = 100,000		& NUM\_OPS = 100			\\
		& MAX\_WAIT = 0.9ms			& MAX\_WAIT = 7.5ms			& MAX\_WAIT = 1ms			\\
\hline
		& 	CONC = 1				&	CONC = 1				& CONC = 10					\\
TS-CAS	& 	SLICE = 20				& 	SLICE = 25				& SLICE = 6					\\
\hline
\end{tabular}
\end{table}

\subsubsection*{Xeon results}
\begin{figure}[bt!]
	\begin{subfigure}{.5\textwidth}\centering
		\includegraphics[scale=0.75]{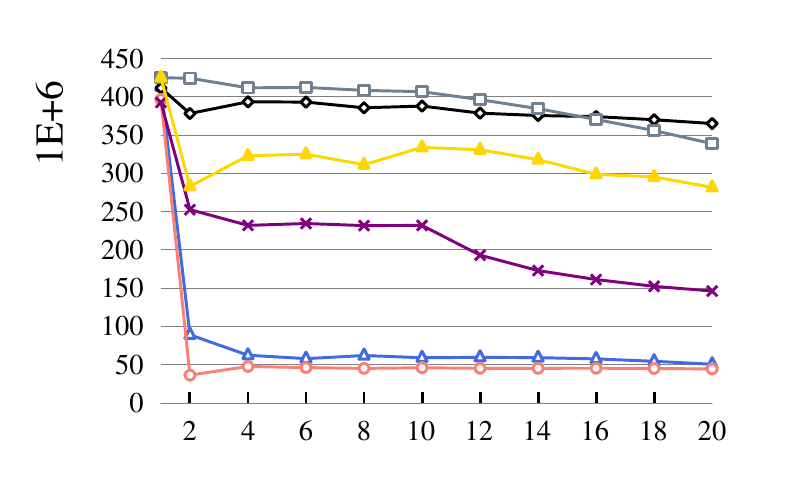}
		\caption{Xeon: Successes}
		\label{figure:XeonJavaCAS}
	\end{subfigure}
	\begin{subfigure}{.5\textwidth}\centering
		\includegraphics[scale=0.75]{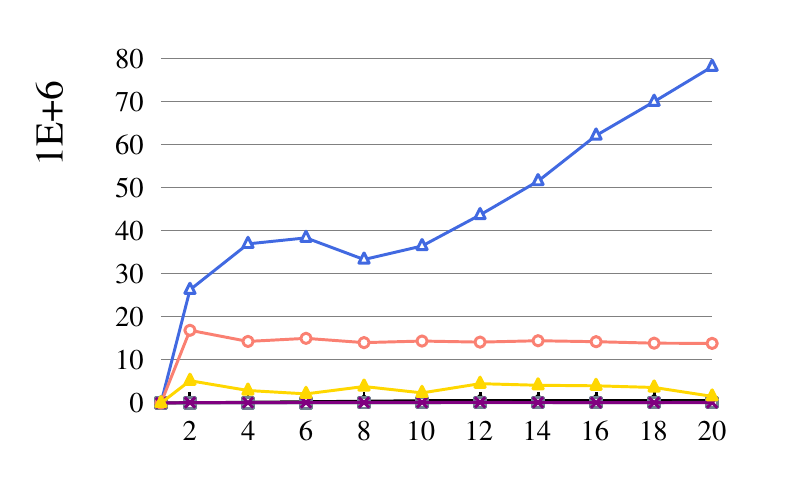}
		\caption{Xeon: Failures}
		\label{figure:XeonJavaFailures}
	\end{subfigure}
	\begin{subfigure}{.5\textwidth}\centering
		\includegraphics[scale=0.65]{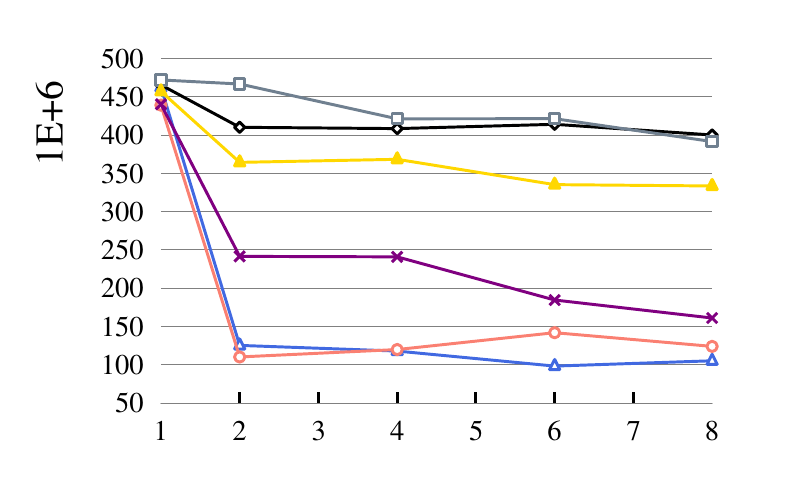}
		\caption{i7: Successes}
		\label{figure:i7JavaCAS}
	\end{subfigure}
	\begin{subfigure}{.4\textwidth}\centering
		\includegraphics[scale=0.6]{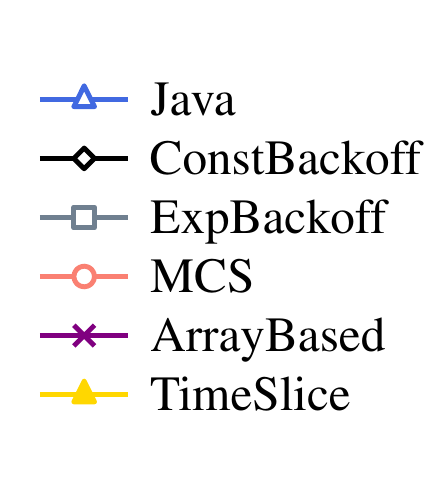}
	\end{subfigure}
	\caption{Xeon \& i7 CAS: Number of successful and failed CAS operations as a function of concurrency level.}
\end{figure}

:

Figure \ref{figure:XeonJavaCAS} shows the throughput (the number of successful CAS operations) on the Xeon machine as a function of the concurrency level. It can be seen that the throughput of Java CAS falls steeply for concurrency levels of 2 or more. Whereas a single thread performs approximately 413M successful CAS operations in the course of the test, the number of successful CAS operations is only approximately 89M for 2 threads and 62M for 4 threads. For higher concurrency levels, the number of successes remains in the range of 50M-59M operations.

In sharp contrast, both the constant wait and exponential backoff CAS algorithms are able to maintain high throughput across the concurrency range. Exponential backoff is slightly better up until 16 threads, but then its throughput declines to below 350M and falls below constant backoff. The throughput of both these algorithms exceeds that of Java CAS by a factor of more than 4 for 2 threads and their performance boost grows to a factor of between 6-7 for higher concurrency levels.

The time slice algorithm is the 3'rd performer in this test, outperforming Java CAS by a factor of between 3-5.6 but providing only between 65\%-87\% the throughput of constant and exponential backoff.

The array based algorithm incurs some overhead and performs only approximately 390M successful operations in the single thread tests. In higher concurrency levels, its throughput exceeds that of Java CAS by a factor of between 2.5-3 but it is consistently outperformed by the simpler backoff algorithms by a wide margin. MCS-CAS is the worst performer on the Xeon CAS benchmark and is outperformed by all other algorithms across the concurrency range.

More insights into these results are provided by Figure \ref{figure:XeonJavaFailures}, which shows the numbers of CAS failures incurred by the algorithms. All algorithms except for MCS-CAS incur orders-of-magnitude less failures than Java CAS. Specifically, for concurrency level 20, Java CAS incurs almost 80M CAS failures, three orders of magnitude more than constant backoff which incurs approximately 569K failures. Exponential backoff incurs approximately 184K failures. Array based incurs approximately 104K failures. MCS-CAS incurs a high number of failures since the tuning of its parameters sets the contention threshold to 8, implying that it is much less likely to enter high contention mode than array based. This high threshold indicates that MCS-CAS is not a good CAS contention management algorithm for Xeon.

\subsubsection*{i7 results}:

Figure \ref{figure:i7JavaCAS} shows the CAS throughput on the i7 machine as a function of the concurrency level. It can be seen that both the absolute and relative performance of the evaluated algorithms are very similar to the behavior on the Xeon machine. The numbers of CAS failures are also very similar to Xeon (for corresponding concurrency levels) and therefore a figure showing these numbers is not provided.

\subsubsection*{SPARC results}
\begin{figure}[bt]
	\begin{subfigure}{\textwidth}
		\includegraphics[scale=0.60]{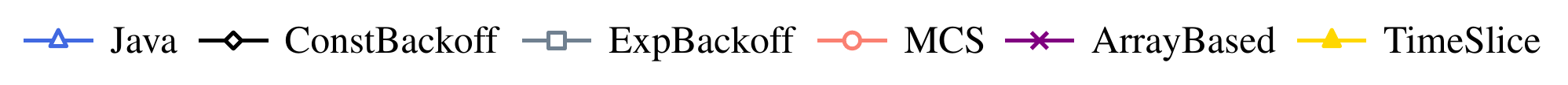}
	\end{subfigure}	
	\begin{subfigure}{.5\textwidth}
		\includegraphics[scale=0.75]{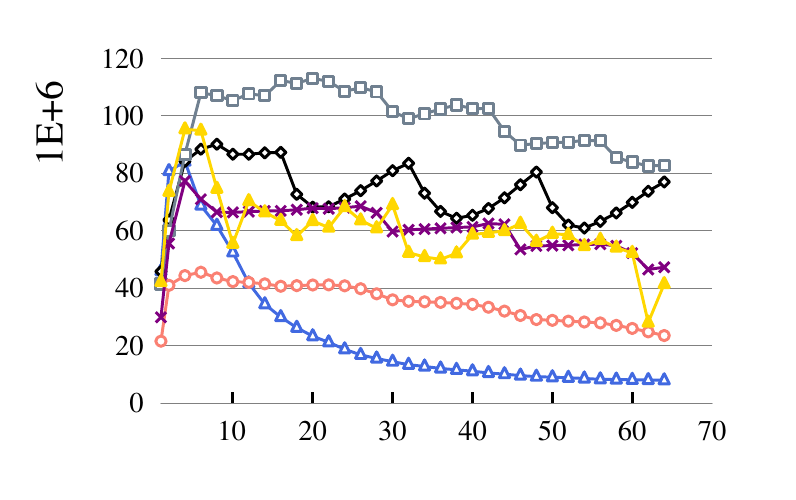}
		\caption{Successes}
		\label{figure:SparcJavaCAS}
	\end{subfigure}
	\begin{subfigure}{.5\textwidth}
		\includegraphics[scale=0.75]{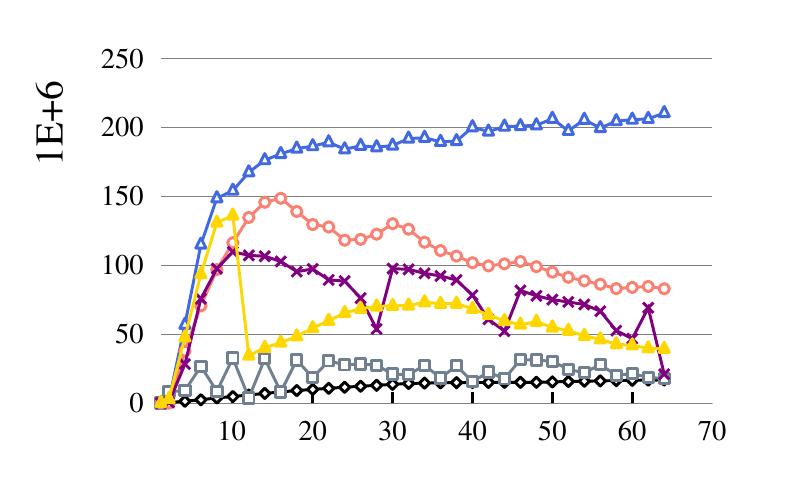}
		\caption{Failures}
		\label{figure:SPARCJavaFailures}
	\end{subfigure}
	\caption{SPARC CAS: Number of successful and failed CAS operations as a function of concurrency level.}
\end{figure}

:

Figure \ref{figure:SparcJavaCAS} shows the throughput of the evaluated algorithms in the CAS benchmark on the SPARC machine. Unlike Xeon where Java CAS does not scale at all, on SPARC the performance of Java CAS scales from 1 to 4 threads but then quickly deteriorates, eventually falling to about 16\% of the single thread performance, less than 9\% of the performance of 4 threads. More specifically, in the single thread test, Java CAS performs slightly more than 48M successful CAS operations and its performance reaches a peak of almost 90M operations at 4 threads. Java CAS is the worst performer for concurrency levels 12 or higher and its throughput drops to about 8M for 64 threads.

The exponential backoff CAS is the clear winner on the SPARC CAS benchmark. Its throughput is slightly lower than that of Java CAS for concurrency levels 1 and 2, but for higher concurrency levels it outperforms Java CAS by a wide margin that grows with concurrency. For concurrency levels 28 or more, exponential backoff completes more than 7 times successful CAS operations and the gap peaks for 54 thread where Java CAS is outperformed by a factor of almost 12.

The constant wait CAS is second best. Since it has smaller overhead than exponential backoff CAS, it slightly outperforms it in the single thread test, but for higher concurrency levels it is outperformed by exponential backoff by a margin of up to 56\%.

The high overhead of MCS-CAS and array based manifests itself in the single thread test, where both provide significantly less throughput than all other algorithms. For higher concurrency levels, both MCS-CAS and array based perform between 20M-60M successful CAS operations, significantly more than Java CAS but much less than the constant and exponential backoff algorithms.

Figure \ref{figure:SPARCJavaFailures} shows the numbers of CAS failures incurred by the algorithms. Constant backoff and exponential bakcoff incur the smallest number of failures, an order of magnitude less failures than Java CAS. Array based, time slice and MCS-CAS incur more failures than the two backoffs, but significantly less than Java CAS in almost all concurrency levels.

Zooming into the numbers of successes and failures incurred by MCS-CAS in low- and high-contention modes, we find that for high concurrency levels, MCS-CAS obtains approximately 10\% of its successes in high-contention mode but also incurs about 10 times more failures in low-contention mode than in high-contention mode.

\subsubsection*{Analysis}:

As shown by Figures \ref{figure:XeonJavaCAS}, \ref{figure:i7JavaCAS} and \ref{figure:SparcJavaCAS}, whereas on the SPARC the number of successes in the CAS benchmark scales up to 4 or 8 threads (depending on the contention management algorithm being used), no such scalability occurs on the Xeon or the i7 platforms. We now explain the architectural reasons for this difference. This requires some background which we now provide.

The SPARC T2+ processor chip contains 8 cores where each core has a private 8KB L1 data cache and 2 pipelines with 4 hardware thread contexts per pipeline, for a total of 64 hardware thread contexts per chip. The L1 data caches, which are physically indexed and physically tagged, use a write-through policy where stores do not allocate. The 8 cores are connected via an intra-chip cross-bar to 8 L2 banks. Based on a hash of the physical address, the cross-bar directs requests to one of the 8 L2 cache banks. The L2 banks are 16-way set associative and have a total capacity of 4MB. Pairs of banks share DRAM channels. All store instructions, including CAS, pass over the cross-bar to the L2 cache. For coherence, the L2, which is inclusive of all L1s, maintains a reverse directory of which L1 instances hold a given line. L1 lines are either valid or invalid; there are no cache-to-cache transfers between L1 caches. T2+ processors enjoy very short cache-coherent communication latencies relative to other processors. On an otherwise unloaded system, a coherence miss can be satisfied from the L2 in under 20 cycles.

CAS instructions are implemented at the interface between the cores and the cross-bar. For ease of implementation, CAS instructions, whether successful or not, invalidate the line from the issuer's L1. A subsequent load from that same address will miss in the L1 and revert to the L2. The cross-bar and L2 have sufficient bandwidth and latency, relative to the speed of the cores, to allow load-CAS benchmarks to scale beyond just one thread, as we see in Figure \ref{figure:SparcJavaCAS}.

We now describe why such scalability is not observed on the XEON and i7 platforms, as seen by Figures \ref{figure:XeonJavaCAS} and \ref{figure:i7JavaCAS}. Modern x86 processors tend to have deeper cache hierarchies, often adding core-local MESI L2 caches connected via an on-chip coherent interconnect fabric and backed by a chip-level L3. Intra-chip inter-core communication is accomplished by L2 cache-to-cache transfers. With respect to coherence, a store instruction is no different than a CAS -- both need to issue request-to-own bus operations, if necessary, to make sure the underlying line can be modified. That is, CAS is performed "locally" in the L1 or L2.

In addition to the cost of obtaining ownership, load-CAS benchmarks may also be subject to a number of confounding factors on x86. As contention increases and the CAS starts to fail more frequently, branch predictors can be trained to expect the failure path, so when the CAS is ultimately successful the thread will incur a branch misprediction penalty. In contrast, T2+ does not have a branch predictor.

Furthermore, some x86 processors have an optimization that allows speculative coherence probes. If a load is followed in close succession, in program order, by a store or CAS to the same address, the processor may need to send coherence request messages to upgrade the line to writable state in its local cache at the time of the load. This avoids the situation where the load induces a read-to-share bus transaction followed in short order by a transaction to upgrade the line to writable state. While useful, under intense communication traffic this facility can cause excessive invalidation. Finally, we note that coherence arbitration for lines is not necessarily fair over the short term, and in turn this can impact performance.

\newpage
\subsubsection*{Fairness}:

\begin{table}
\caption{Fairness measures.}
\label{table:summary-fairness}
\begin{center}
\begin{tabular}{|l|c|c|c|c|}
\hline
		&	\multicolumn{2}{l|}{Normal stdev}	&	\multicolumn{2}{l|}{Jain's Index}	\\
		&	Xeon	&	SPARC	&	Xeon	&	SPARC	\\
\hline
\hline
Java	&	0.291	&	0.164 	&	0.900 	&	0.961	\\
\hline
CB-CAS	&	0.077	& 	0.196 	&	0.992 	&	0.957	\\
\hline
EXP-CAS	&	0.536	& 	0.936 	&	0.761	 &	0.588	\\
\hline
MCS-CAS	&	0.975	& 	0.596 	&	0.563	 &	0.727	\\
\hline
AB-CAS	&	0.001	& 	0.822	&	1.000	 &	0.638	\\
\hline
TS-CAS	&	0.829	& 	0.211 	&	0.605	 &	0.946	\\
\hline
\end{tabular}
\end{center}
\end{table}

Table \ref{table:summary-fairness} summarizes the fairness measures of the synthetic CAS benchmarks. We used normalized standard deviation and Jain's fairness index to quantify the fairness of individual threads' throughput for each concurrency level, and then took the average over all concurrency levels. The widely used Jain's index for a set of $n$ samples is the quotient of the square of the sum and the product of the sum of squares by $n$. Its value ranges between $1/n$ (lowest fairness) and $1$ (highest fairness). It equals $k/n$ when $k$ threads have the same throughput, and the other $ n-k $ threads are starved. We see that CB-CAS and TS-CAS provide comparable and even superior fairness to Java CAS while the rest of the algorithms provide less fairness.

\subsection{FIFO queue}

To further investigate the impact of our CAS contention management algorithms, we experimented with the FIFO queue algorithm of Michael and Scott \cite{DBLP:conf/podc/MichaelS96} (MS-queue). We used the Java code provided in Herlihy and Shavit's book \cite{Herlihy:2008:AMP:1734069} without any optimizations. The queue is represented by a list of nodes and by \emph{head} and \emph{tail} atomic references to the first and last entries in the list, which become hot spots under high contention.

We evaluated four versions of the MS-queue: one using Java's AtomicReference objects (called J-MSQ), and the other three replacing them by ConstantBackoff\-CAS, ExpBackoffCAS and TimeSliceCAS objects (respectively called CB-MSQ, Exp-MSQ and TS-MSQ). MCS and array based were consisten\-tly outperformed and are therefore omitted from the following comparison. We compared these algorithms with the Java 6 ConcurrentLinkedQueue class from the java.util.concurrent package,\footnote{We used a slightly modified version in which direct usage of Java's \texttt{Unsafe} class was replaced by an AtomicReference mediator.} and the flat-combining queue algorithm \cite{FlatCombining}.\footnote{We used the Java implementation provided by Tel-Aviv University's Multicore Computing Group.}

The ConcurrentLinkedQueue class, written by Doug Lea, implements an algorithm (henceforth simply called Java 6 queue) that is also based on Michael and Scott's algorithm. However, the Java 6 queue algorithm incorporates several significant optimizations such as performing lagged updates of the head and tail references and using lazySets instead of normal writes.

We conducted the following test. For varying number of threads, each thread repeatedly performed either an \texttt{enqueue} or a \texttt{dequeue} operation on the data structure for a period of 5 seconds. The queue is pre-populated by 1000 items. A pseudo-random sequence of 128 integers is generated by each thread independently before the test starts where the $i$'th operation of thread $t$ is an \texttt{enqueue} operation if integer $(i \mod 128)$ is even and is a \texttt{dequeue} operation otherwise. Each thread counts the number of operations it completes on the queue. These local counters are summed up at the end of the test. Each data point is the average of 10 independent runs. In order to make the results comparable between the different platforms, the same set of 10 pre-generated seeds was used to initialize the random generator.

\begin{figure}[!t]
	\begin{subfigure}{.5\textwidth}\centering
		\includegraphics[scale=0.75]{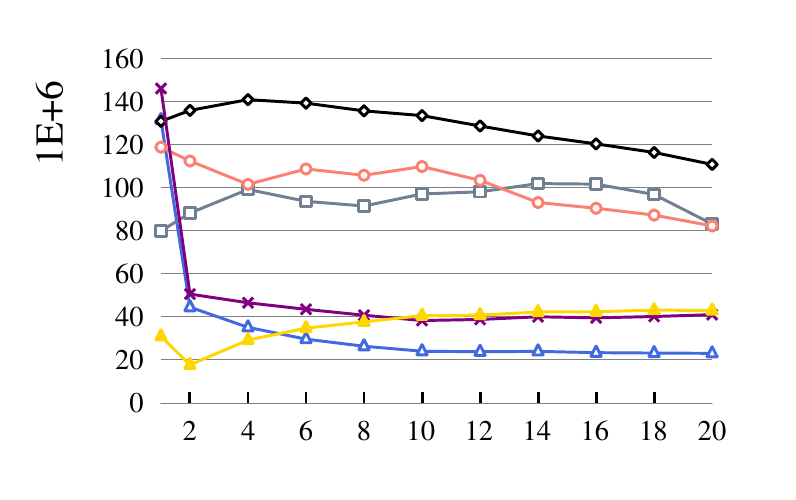}
		\caption{Xeon}
		\label{figure:XeonJavaQueue}
	\end{subfigure}
	\begin{subfigure}{.5\textwidth}\centering
	\includegraphics[scale=0.75]{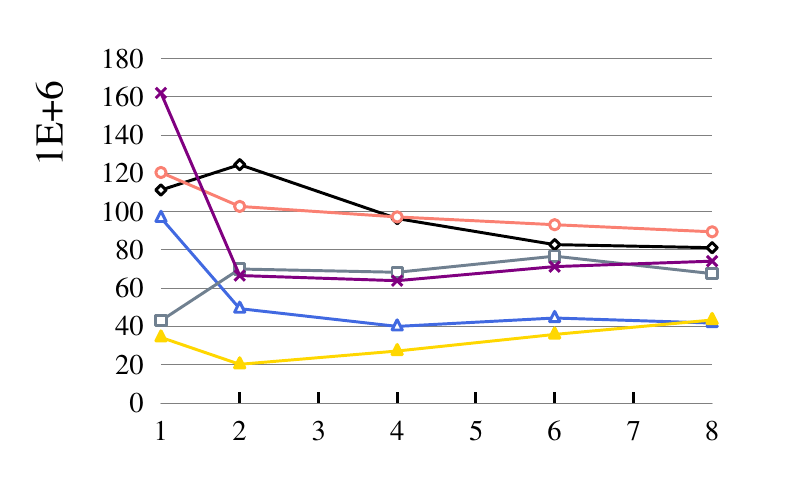}
	\caption{i7}
	\label{figure:i7JavaQueue}
	\end{subfigure}
	\begin{subfigure}{.65\textwidth}
		\includegraphics[scale=0.9]{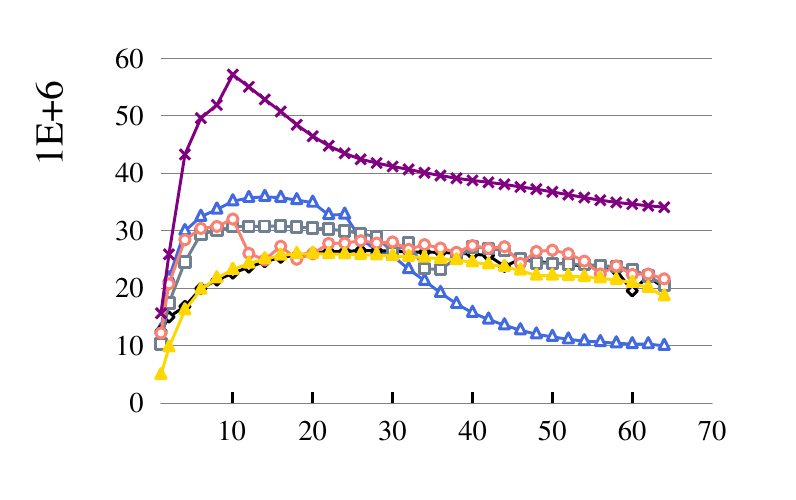}
		\caption{SPARC}
		\label{figure:SparcJavaQueue}
	\end{subfigure}
	\begin{subfigure}{.25\textwidth}
		\includegraphics[scale=0.6]{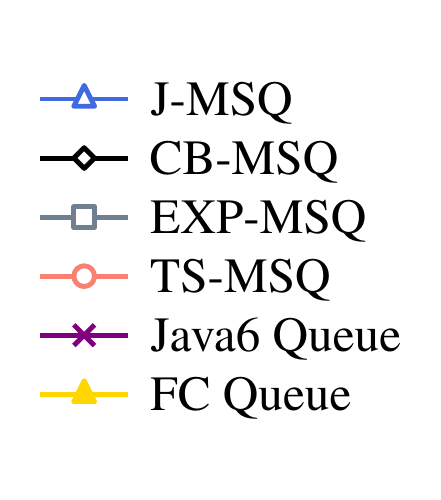}
	\end{subfigure}

	\caption{Queue: Number of completed ops as a function concurrency level.}
\end{figure}

Figures \ref{figure:XeonJavaQueue}-\ref{figure:SparcJavaQueue} show the results of the queue tests on the platforms on which we ran our experiments, using the optimal parameter values of Table \ref{table:summary-tuned-parameters}.

\subsubsection*{Xeon results}:

As shown by Figure \ref{figure:XeonJavaQueue}, CB-MSQ is the best queue implementation, outperforming the Java-CAS based queue in all concurrency levels by a factor of up to 6 (for 16 threads).

Surprisingly, CB-MSQ also outperforms the Java 6 queue by a wide margin in all concurrency levels except 1, in spite of the optimizations incorporated to the latter. More specifically, in the single thread test, the performance of the Java 6 queue exceeds that of CB-MSQ by approximately 15\%. In higher concurrency levels, however, CB-MSQ outperforms Java 6 queue by a factor of up to 3.5. Java 6 queue is outperformed in all concurrency levels higher than 1 also by EXP-MSQ and TS-MSQ. The FC queue hardly scales on this test and is outperformed by almost all algorithms in most concurrency levels.

However, whereas in the Xeon CAS benchmark the constant backoff and exponential backoff provided nearly the same throughput, in the queue test CB-MSQ outperforms EXP-MSQ by a wide margin in most concurrency levels.

J-MSQ has the worst performance in all concurrency levels higher than 1. It is outperformed by CB-MSQ by a factor of between 2-6 in all concurrency levels higher than 1.

\subsubsection*{i7 results}:

Figure \ref{figure:i7JavaQueue} shows the results of the queue test on the i7 machine. The differences between the algorithms in this test are less significant than on the Xeon. CB-MSQ and TS-MSQ provide the highest throughput for all concurrency levels except for 1. CB-MSQ peaks at 2 threads providing 124.5M operations, after which it starts to decline until reaching 81M for 8 threads. TS-MSQ maintains a consistent throughput of 90M-100M for concurrency levels higher than 1 and is the best performer for concurrency levels 6 or more. EXP-MSQ, which was significantly better than the Java 6 queue on Xeon, outperforms it only by roughly 5\% in this test for concurrency levels of 2-6, and by 9\% for 8 threads.

J-MSQ falls from 96.7M for 1 thread to about 40M-44M for higher concurrency levels, exhibiting similar behavior to the Xeon test. FC queue hardly scales in this test as well, providing the lowest throughput in all concurrency levels. TS-MSQ outperforms J-MSQ by factor of between 2.1-2.4 for all concurrency levels except for 1.



\subsubsection*{SPARC results}:

Figure \ref{figure:SparcJavaQueue} shows the results of the queue test on the SPARC machine. Here, unlike on Xeon and i7, the Java 6 queue has the best throughput in all concurrency levels, outperforming TS-MSQ - which is second best in most concurrency levels - by a factor of up to 2. It seems that the optimizations of the Java 6 algorithm are more effective on the SPARC architecture. CB-MSQ starts low but its performance scales up to 30 threads where it slightly exceeds that of EXP-MSQ.

J-MSQ scales up to 14 threads where it performs approximately 36M queue operations, but quickly deteriorates in higher concurrency levels and its throughput falls to less than 10M operations with 64 threads. This is similar to the decline exhibited by Java CAS in the CAS benchmark, except that the graph is ``stretched'' and the decline is slightly milder. The reason for this change is that the effective levels of CAS contention on the data-structure's variables are reduced in the queue implementations, since the code of the MS-queue algorithm contains operations other than CAS. For concurrency levels 40 or higher, J-MSQ is outperformed by EXP-MSQ by a factor of up to 2.4 (for 54 threads). Unlike on Xeon, the FC queue scales on SPARC up to 20 threads, when its performance almost equals that of the simple backoff schemes.


\subsection{Stack}


We also experimented with the lock-free stack algorithm of Treiber.\footnote{The first non-blocking implementation of a concurrent list-based stack appeared in the IBM System 370 principles of operation manual in 1983 \cite{IBM370-manual} and used the double-width compare-and-swap (CAS) primitive. Treiber's algorithm is a variant of IBM's algorithm, in which push operations use a single-word-width CAS instead of double-width compare-and-swap.} The stack is represented by a list of nodes and a reference to the top-most node is stored by an AtomicReference object.

We evaluated five versions of the Treiber algorithm: one using Java's AtomicReference objects (called J-Treiber), and the other three replacing them by the ConstantBackoffCAS, ExpBackoffCAS and TimeSliceCAS (respectively called CB-Treiber, Exp-Treiber and TS-Treiber). We also compared with a Java implementation of the elimination-backoff stack (EB stack) of Hendler et al. \cite{DBLP:journals/jpdc/HendlerSY10}.\footnote{We used IBM's implementation available from the Amino Concurrent Building Blocks project at http://amino-cbbs.wiki.sourceforge.net/} The elimination-backoff stack copes with high-contention by attempting to pair-up concurrent push and pop operations that ``collide'' on entries of a so-called \emph{elimination array}. In addition, it employs an exponential-backoff scheme after a CAS failure.

The structure of the Stack test is identical to that of the Queue test: each thread repeatedly performs either a \texttt{push} or a \texttt{pop} operation on the stack for a period of 5 seconds. The stack is pre-populated by 1000 items. A pseudo-random sequence of 128 bits is generated by each thread independently before the test starts where the $i$'th operation of thread $t$ is an \texttt{push} operation if bit $(i \mod 128)$ is true and is a \texttt{pop} operation otherwise. Each data point is the average of 10 independent runs.

\begin{figure}[!t]
	\begin{subfigure}{.5\textwidth}\centering
		\includegraphics[scale=0.75]{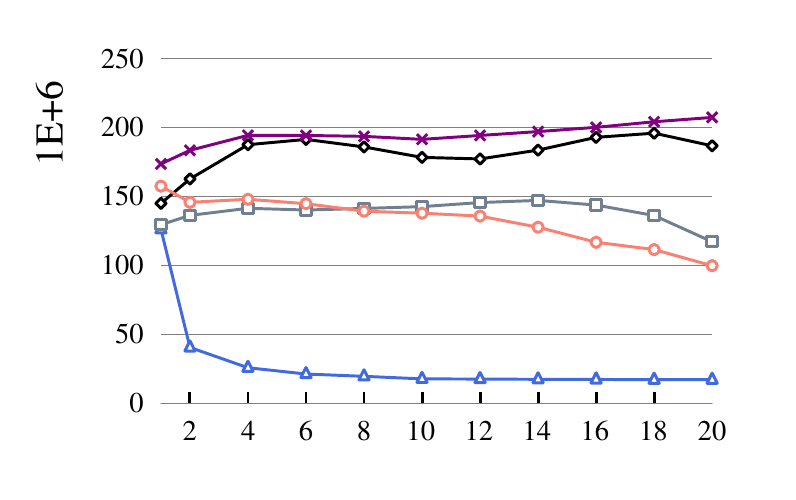}
		\caption{Xeon}
		\label{figure:XeonJavaStack}
	\end{subfigure}
	\begin{subfigure}{.5\textwidth}\centering
		\includegraphics[scale=0.75]{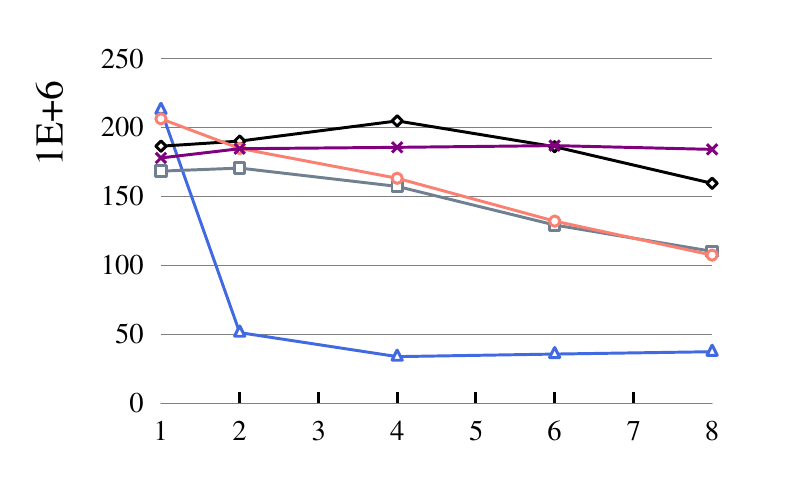}
		\caption{i7}
		\label{figure:i7JavaStack}
	\end{subfigure}
	\begin{subfigure}{.65\textwidth}
		\includegraphics[scale=0.9]{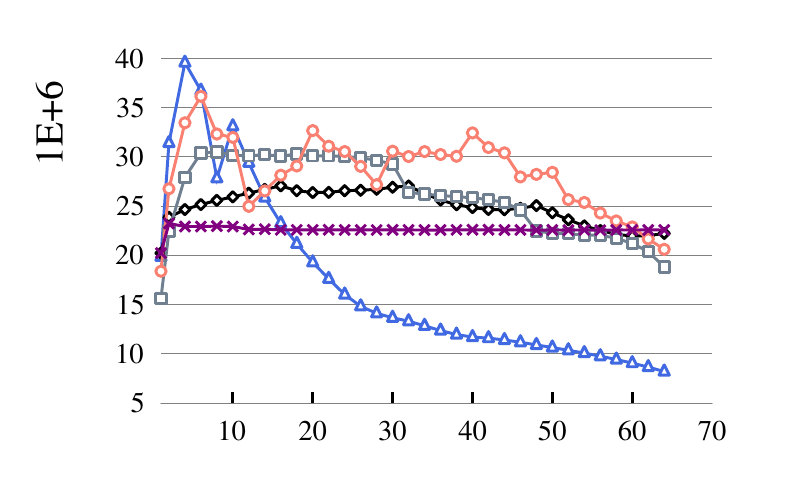}
		\caption{SPARC}
		\label{figure:SparcJavaStack}
	\end{subfigure}
	\begin{subfigure}{.25\textwidth}
		\includegraphics[scale=0.6]{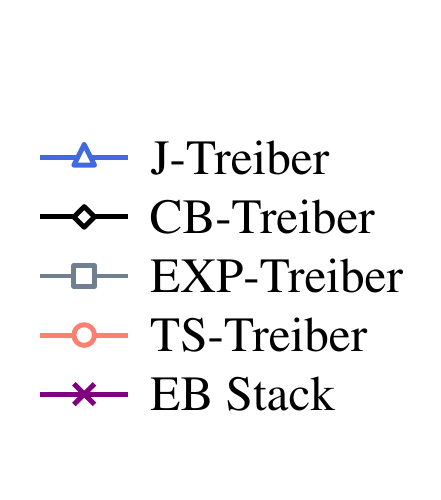}
	\end{subfigure}

	\caption{Stack: Number of completed ops as a function concurrency level.}
\end{figure}

Figures \ref{figure:XeonJavaStack}-\ref{figure:SparcJavaStack} show the results of the stack tests on the three platforms, using the optimal parameter values of Table \ref{table:summary-tuned-parameters}.

\subsubsection*{Xeon results}:

Figure \ref{figure:XeonJavaStack} shows the results of the stack test on Xeon. As with all Xeon test results, also in the stack test, the implementation using Java's AtomicReference suffers from a steep performance decrease as concurrency levels increase, falling from throughput of approximately 126M stack operations in the single thread test to about 17M operations for 20 threads, approximately 13\% of the single thread performance.

The EB stack is the winner of the Xeon stack test and CB-Treiber is second-best lagging behind only slightly. CB-Treiber maintains and even exceeds its high single-thread throughput across the concurrency range, scaling up from 144M operations for a single thread to 195M operations for 18 threads, outperforming J-Treiber by a factor of 11.5 for 18 threads. EXP-Treiber and TS-Treiber are second best, with performance lagging behind CB-Treiber in all concurrency levels by between 20\%-40\%.


\subsubsection*{i7 results}:

Figure \ref{figure:i7JavaStack} shows the results of our evaluation on the i7. The EB stack and CB-Treiber algorithms are the best performers. CB-Treiber has the upper hand in low concurrency levels, providing between 5\%-10\% more throughput than EB stack for 1-4 threads. It scales up to 4 threads, then starts to deteriorate and levels up with EB stack at 6 threads, where the throughput of both algorithms is approximately 186M. EB Stack maintains a consistent throughput of about 185M through all concurrency levels, outperforming CB-Treiber at 8 threads by 15\%.

EXP-Treiber is significantly outperformed by both EB stack and CB-Treiber. Its throughput declines from about 168M in the single-thread test to approximately 110M for 8 threads. TS-Treiber starts high at 206M for 1 thread, but deteriorates to 107M, in high correlation with Exp-Treiber. J-Treiber exhibits a  curve similar to the corresponding Java CAS in the CAS benchmark; it falls from more than 212M for a single thread to only 37M for 8 threads, and is outperformed by CB-Treiber in all concurrency levels except for 1 by a wide margin of up to 6.2.


\subsubsection*{SPARC results}:

Figure \ref{figure:SparcJavaStack} shows the results of the stack tests on SPARC. J-Treiber scales up to 6 threads where it reaches its peak performance of 39.5M stack operations. Then its performance deteriorates with concurrency and reaches less than 10M operations for 64 threads. From concurrency level 18 and higher, J-Treiber has the lowest throughput. TS-Treiber has the highest throughput in most medium and high concurrency levels, with EXP-Treiber mostly second best. Unlike on XEON, EB stack is almost consistently and significantly outperformed on SPARC by all simple backoff algorithms.

TS-Treiber has the highest throughput for 30 threads or more (with the exception of concurrency levels 62-64) and outperforms J-Treiber in high concurrency levels by a factor of up to 3. CB-Treiber starts low but scales up to 18 where it levels up at about 27M until it starts to deteriorate at 34 threads and higher, matching and even slightly exceeding TS-Treiber and EXP-Treiber above 62 threads.
\section{Discussion}
\label{sec:discussion}

We conduct what is, to the best of our knowledge, the first study on the impact of contention management algorithms on the efficiency of the CAS operation. We implemented several Java classes that encapsulate calls to Java's AtomicReference class by CAS contention management algorithms. We then evaluated the benefits gained by these algorithms on the Xeon, SPARC and i7 platforms by using both a synthetic benchmark and CAS-based concurrent data-structure implementations of stacks and queues.

Out of the contention management approaches we have experimented with, the three simplest algorithms - constant backoff, exponential backoff and time-slice - yielded the best results, primarily because they have very small overheads. The more complicated approaches - the MCS-CAS and array-based algorithms - provided better results than direct calls to AtomicReference in most tests, but were significantly outperformed by the simpler algorithms.

\emph{Our evaluation demonstrates that encapsulating Java's AtomicReference by classes that implement lightweight contention management support can improve the performance of CAS-based algorithms considerably. } We also compared relatively simple data-structure implementations that use our CAS contention management classes with more complex implementations that employ data-structure specific optimizations and use AtomicReference objects.

\emph{We have found that, in some cases, simpler and non-optimized data-structure implementations that apply efficient contention management for CAS operations yield better performance than that of highly optimized implementations of the same data-structure that use Java's AtomicReference directly.}

Our results imply that encapsulating invocations of CAS by lightweight contention management classes is a simple and generic way of improving the performance of concurrent objects.

This work may be extended in several directions. First, we may have overlooked CAS contention management algorithms that yield better results. Second, our methodology tuned the platform-dependent parameters of contention management algorithms by using the CAS benchmark. Although the generality of this approach is appealing, tuning these parameters per data-structure may yield better results. Moreover, a dynamic tuning may provide a general, cross data-structure, cross CPU, solution.

It would also be interesting to investigate if and how similar approaches can be used for other atomic-operation related classes in both Java and other programming languages such as C++.

Finally, combining contention management algorithms at the atomic operation level with optimizations at the data-structure algorithmic level may yield more performance gains than applying only one of these approaches separately. We leave these research directions for future work.

\bibliographystyle{plain}
\bibliography{paper}

\begin{thebibliography}{10}

\bibitem{DBLP:conf/wdag/AfekHM11}
Yehuda Afek, Michael Hakimi, and Adam Morrison.
\newblock Fast and scalable rendezvousing.
\newblock In {\em DISC}, pages 16--31, 2011.

\bibitem{DBLP:conf/europar/AfekKNS10}
Yehuda Afek, Guy Korland, Maria Natanzon, and Nir Shavit.
\newblock Scalable producer-consumer pools based on elimination-diffraction
  trees.
\newblock In {\em Euro-Par (2)}, pages 151--162, 2010.

\bibitem{AW98}
Hagit Attiya and Jennifer Welch.
\newblock {\em Distributed Computing: Fundamentals, Simulations and Advanced
  Topics (2nd edition)}.
\newblock John Wiley Interscience, March 2004.

\bibitem{DBLP:conf/wdag/BasinFKKP11}
Dmitry Basin, Rui Fan, Idit Keidar, Ofer Kiselov, and Dmitri Perelman.
\newblock Caf{\'e}: Scalable task pools with adjustable fairness and
  contention.
\newblock In {\em DISC}, pages 475--488, 2011.

\bibitem{Boyd-Wickizer-Locks}
Silas Boyd-Wickizer, M.~Frans Kaashoek, Robert Morris, and Nickolai Zeldovich.
\newblock Non-scalable locks are dangerous.
\newblock In {\em Proceedings of the Linux Symposium}, pages 119--130, 2012.

\bibitem{Itanium2001}
Intel Corporation.
\newblock {\em Intel Itanium Architecture Software Developer's Manual}.
\newblock 2006.

\bibitem{DBLP:conf/ppopp/FatourouK12}
Panagiota Fatourou and Nikolaos~D. Kallimanis.
\newblock Revisiting the combining synchronization technique.
\newblock In {\em PPOPP}, pages 257--266, 2012.

\bibitem{DBLP:conf/opodis/GidenstamST10}
Anders Gidenstam, H{\aa}kan Sundell, and Philippas Tsigas.
\newblock Cache-aware lock-free queues for multiple producers/consumers and
  weak memory consistency.
\newblock In {\em OPODIS}, pages 302--317, 2010.

\bibitem{DBLP:conf/spaa/GidronKPP12}
Elad Gidron, Idit Keidar, Dmitri Perelman, and Yonathan Perez.
\newblock Salsa: scalable and low synchronization numa-aware algorithm for
  producer-consumer pools.
\newblock In {\em SPAA}, pages 151--160, 2012.

\bibitem{DBLP:conf/sc/GoodmanLJ11}
Eric~L. Goodman, M.~Nicole Lemaster, and Edward Jimenez.
\newblock Scalable hashing for shared memory supercomputers.
\newblock In {\em SC}, page~41, 2011.

\bibitem{FlatCombining}
Danny Hendler, Itai Incze, Nir Shavit, and Moran Tzafrir.
\newblock Flat combining and the synchronization-parallelism tradeoff.
\newblock In {\em SPAA}, pages 355--364, 2010.

\bibitem{DBLP:conf/wdag/HendlerIST10}
Danny Hendler, Itai Incze, Nir Shavit, and Moran Tzafrir.
\newblock Scalable flat-combining based synchronous queues.
\newblock In {\em DISC}, pages 79--93, 2010.

\bibitem{DBLP:journals/jpdc/HendlerSY10}
Danny Hendler, Nir Shavit, and Lena Yerushalmi.
\newblock A scalable lock-free stack algorithm.
\newblock {\em J. Parallel Distrib. Comput.}, 70(1):1--12, 2010.

\bibitem{Herlihy:1993:MIH:161468.161469}
Maurice Herlihy.
\newblock A methodology for implementing highly concurrent data objects.
\newblock {\em ACM Trans. Program. Lang. Syst.}, 15(5):745--770, November 1993.

\bibitem{DBLP:journals/tocs/HerlihyLS95}
Maurice Herlihy, Beng-Hong Lim, and Nir Shavit.
\newblock Scalable concurrent counting.
\newblock {\em ACM Trans. Comput. Syst.}, 13(4):343--364, 1995.

\bibitem{Herlihy:1993:TMA:165123.165164}
Maurice Herlihy and J.~Eliot~B. Moss.
\newblock Transactional memory: architectural support for lock-free data
  structures.
\newblock In {\em Proceedings of the 20th annual international symposium on
  computer architecture}, ISCA '93, pages 289--300, New York, NY, USA, 1993.
  ACM.

\bibitem{Herlihy:2008:AMP:1734069}
Maurice Herlihy and Nir Shavit.
\newblock {\em The Art of Multiprocessor Programming}.
\newblock Morgan Kaufmann Publishers Inc., San Francisco, CA, USA, 2008.

\bibitem{DBLP:conf/wdag/HerlihyST08}
Maurice Herlihy, Nir Shavit, and Moran Tzafrir.
\newblock Hopscotch hashing.
\newblock In {\em DISC}, pages 350--364, 2008.

\bibitem{DBLP:journals/dc/HerlihySW96}
Maurice Herlihy, Nir Shavit, and Orli Waarts.
\newblock Linearizable counting networks.
\newblock {\em Distributed Computing}, 9(4):193--203, 1996.

\bibitem{waitfree}
M.P. Herlihy.
\newblock Wait-free synchronization.
\newblock {\em ACM Transactions On Programming Languages and Systems},
  13(1):123--149, January 1991.

\bibitem{IBM370-manual}
IBM.
\newblock {\em IBM System/370 Extended Architecture, Principles of Operation,
  publication no. SA22-7085}.
\newblock 1983.

\bibitem{DBLP:journals/cacm/SchererLS09}
William N.~Scherer III, Doug Lea, and Michael~L. Scott.
\newblock Scalable synchronous queues.
\newblock {\em Commun. ACM}, 52(5):100--111, 2009.

\bibitem{DBLP:journals/dc/Ladan-MozesS08}
Edya Ladan-Mozes and Nir Shavit.
\newblock An optimistic approach to lock-free fifo queues.
\newblock {\em Distributed Computing}, 20(5):323--341, 2008.

\bibitem{mcs}
J.~M. Mellor-Crummey and M.~L. Scott.
\newblock Algorithms for scalable synchronization on shared-memory
  multiprocessors.
\newblock {\em ACM Transactions on Computer Systems (TOCS)}, 9(1):21--65, 1991.

\bibitem{DBLP:conf/podc/MichaelS96}
Maged~M. Michael and Michael~L. Scott.
\newblock Simple, fast, and practical non-blocking and blocking concurrent
  queue algorithms.
\newblock In {\em PODC}, pages 267--275, 1996.

\bibitem{Michael:1996:SFP:248052.248106}
Maged~M. Michael and Michael~L. Scott.
\newblock Simple, fast, and practical non-blocking and blocking concurrent
  queue algorithms.
\newblock In {\em Proceedings of the fifteenth annual ACM symposium on
  Principles of distributed computing}, PODC '96, pages 267--275, New York, NY,
  USA, 1996. ACM.

\bibitem{Sparc}
Sun Microsystems.
\newblock {\em UltraSPARC Architecture 2005, Draft D0.9.2}.
\newblock 2008.

\bibitem{DBLP:conf/spaa/MoirNSS05}
Mark Moir, Daniel Nussbaum, Ori Shalev, and Nir Shavit.
\newblock Using elimination to implement scalable and lock-free fifo queues.
\newblock In {\em SPAA}, pages 253--262, 2005.

\bibitem{Motorola86}
Motorola.
\newblock {\em {MC68000} Programmer's Reference Manual}.
\newblock 1992.

\bibitem{Scherer:2005:ACM:1073814.1073861}
William~N. Scherer, III and Michael~L. Scott.
\newblock Advanced contention management for dynamic software transactional
  memory.
\newblock In {\em Proceedings of the twenty-fourth annual ACM symposium on
  Principles of distributed computing}, PODC '05, pages 240--248, New York, NY,
  USA, 2005. ACM.

\bibitem{DBLP:journals/jacm/ShalevS06}
Ori Shalev and Nir Shavit.
\newblock Split-ordered lists: Lock-free extensible hash tables.
\newblock {\em J. ACM}, 53(3):379--405, 2006.

\bibitem{DBLP:journals/sigops/TriplettMW10}
Josh Triplett, Paul~E. McKenney, and Jonathan Walpole.
\newblock Scalable concurrent hash tables via relativistic programming.
\newblock {\em Operating Systems Review}, 44(3):102--109, 2010.

\end{thebibliography}

\newpage
\section{Appendix A: the MSC-CAS algorithm}

\begin{algorithm}[H]
\scriptsize
\caption{\label{alg:MCS-CAS}The \texttt{MCS-CAS} class.}

\Public \Class \textbf{MCS-CAS$<$V$>$} \Extends AtomicReference$<$V$>$\;

\BlankLine

\Private \Class \textbf{ThreadRecord} \Proc{
	\Long \emph{modeCount}\;
	\Bool \emph{contentionMode}\;
	\Int \emph{next} = NONE\;
	\Volatile \Bool \emph{notify}\;
}

\BlankLine

\Private ThreadRecord[] \emph{tRecords} = \New ThreadRecord[MAX\_THREADS]\;
\Private AtmicInteger \emph{tail} = \New AtomicInteger(NONE)\;

\BlankLine

%

\Public V \textbf{read}() \Proc{
    ThreadRecord r = \emph{tRecords}[TInd]\;
	\If{r.\emph{contentionMode}\nllabel{MCS-CAS:ifHighContention}}
    {
	    r.\emph{next} = NONE \nllabel{MCS-CAS:setNextToNone}\;
    	\Int \emph{pred} = \emph{tail}.getAndSet(TInd) \nllabel{MCS-CAS:swapTail}\;

	    \If{pred \emph{!= NONE} \nllabel{MCS-CAS:ifPred}}
	    {
            \emph{tRecords}[pred].\emph{next}.set(TInd)
                    \nllabel{MCS-CAS:updatePred}\;
	        r.\emph{notify}.set(\False)\nllabel{MCS-CAS:resetNotify}\;
	        \Long \emph{wait} = MAX\_WAIT \nllabel{MCS-CAS:setWaitingIter}\;
	        \lWhile{$\lnot$r.notify\emph{[}TInd\emph{]} $\land$ \emph{(}wait $>$ 0\emph{)}
	        	\nllabel{MCS-CAS:whileNotNotified}}
			{
	            \emph{wait}=\emph{wait}-1\;
	            \nllabel{MCS-CAS:whileNotNotifiedEnd}
			}
		}
    }
	\Return get() \nllabel{MCS-CAS:lastReadReturn}\;
}

\BlankLine

\Public \Bool \textbf{CAS}(V old, V new) \Proc{
    \Bool ret = compareAndSet(old,new)\nllabel{MCS-CAS:applyCAS}\;
    ThreadRecord r = \emph{tRecords}[TInd]\;
    \uIf {\emph{r}.contentionMode}
    {
        \uIf{\emph{r}.next == \emph{NONE}\nllabel{MCS-CAS:successorWrote}}
        {
            \If{$\lnot$tail.compareAndSet(\emph{TInd}, \emph{NONE})
                    \nllabel{MCS-CAS:swapBack}}
            {
                \Long \emph{wait} = MAX\_WAIT\nllabel{MCS-CAS:waitSucStart}\;
                \lWhile {\emph{r}.next == \emph{NONE} $\land$ \emph{(}wait $>$ 0\emph{)}}
                {
                    \emph{wait}=\emph{wait}-1\nllabel{MCS-CAS:waitSucEnd}\;
                }
                \Int successor = r.\emph{next}\nllabel{MCS-CAS:checkSuccessor}\;
                \lIf{successor $\neq$ NONE}
                {
                    tRecords[successor].notify $=$\True
                        \nllabel{MCS-CAS:signalSuccessor2}\;
                }
            }
            \Else
            {
                \Int successor = r.\emph{next}[TInd]\nllabel{MCS-CAS:readSuccessor}\;
                \emph{tRecords}[successor].\emph{notify} $=$\True
                    \nllabel{MCS-CAS:signalSuccessor}\;
            }

        }

        r.\emph{modeCount} = r.\emph{modeCount} + 1\nllabel{MCS-CAS:incModeCounter}\;
        \If{\emph{r}.modeCount=\emph{NUM\_OPS}}
        {
            r.\emph{modeCount} = 0, r.\emph{contentionMode} = \False
                \nllabel{MCS-CAS:shiftToLowContentionMode}
        }
    }
    \ElseIf{\emph{ret}}
    {
        r.\emph{modeCount} = 0 \nllabel{MCS-CAS:resetModeCount}\;
    }
    \Else
    { \nllabel{MCS-CAS:shiftToContentionModeStart}
        r.\emph{modeCount} = r.\emph{modeCount} + 1\;
        \If{\emph{r}.modeCount == CONTENTION\_THRESHOLD}
        {
            r.\emph{contentionMode} = \True\;
            r.\emph{modeCount} = 0 \nllabel{MCS-CAS:shiftToContentionModeEnd}
        }
    }

    \BlankLine
    \Return ret \nllabel{MCS-CAS:return}\;
}
\end{algorithm}

\newpage

Algorithm \ref{alg:MCS-CAS} presents the \texttt{MCS-CAS} class, which implements this algorithm. Each class instance contains the following two fields. The \emph{tail} field is an atomic integer storing the TInd of the thread that is at the tail of the queue of threads that are currently in high-contention mode. The \emph{tRecords} field is an array of per-thread data records, storing the following fields.\footnote{To cope with false sharing the records are padded with dummy fields (which we ensure that are not optimized-out).} The \emph{contentionMode} field is a boolean, indicating whether the respective thread is in high-contention mode (if true) or in low-contention mode (if false). The \emph{next} field of the record corresponding to $ t $ is an atomic integer, used by $t$'s successor in the queue for communicating its TInd to $t$. The \emph{notify} field of the record corresponding to $ t $ is an atomic integer array used by $t$'s predecessor to signal $t$ when it is allowed to proceed with its read operation. The \emph{modeCount} field is used by a thread to determine when it should shift from high-contention mode to low-contention mode or vice versa as we soon explain.

We start by describing the \texttt{read} operation. If thread $t$ is in low-contention mode, then it simply delegates to the \texttt{get} method of the AtomicReference object to return the current reference value in \lref{MCS-CAS:lastReadReturn}.

If $t$ is in high-contention mode, then it initializes its \emph{next} entry (\lref{MCS-CAS:setNextToNone}) and swaps the value of \emph{tail} to its TInd (\lref{MCS-CAS:swapTail}). After the swap $t$ checks if it has a predecessor (\lref{MCS-CAS:ifPred}). If it does, $t$ writes its TInd to the \emph{next} field of the \emph{pred} entry of its predecessor in the \texttt{tRecords} array, and initializes its \emph{notify} field (\llref{MCS-CAS:updatePred}{MCS-CAS:resetNotify}).

Thread $t$ then waits until it is either notified by its predecessor that it can go ahead or until a platform-dependent waiting time elapses (\llref{MCS-CAS:setWaitingIter}{MCS-CAS:whileNotNotifiedEnd}) and then returns the current reference value.

\noindent Regardless of whether $t$ is in high- or low-contention mode it always leaves the function by returning the CAS success/failure indication in \lref{MCS-CAS:return}. We now describe the \texttt{CAS} operation.

First, the \texttt{compareAndSet} method on the AtomicReference superclass is called, passing to it the \emph{old} and \emph{new} operands (\lref{MCS-CAS:applyCAS}). If $t$ is in high-contention mode then it checks whether a successor has written its TInd to $t$'s \emph{next} field (\lref{MCS-CAS:successorWrote}) and if so signals that successor that it may stop waiting for $t$ (\llref{MCS-CAS:readSuccessor}{MCS-CAS:signalSuccessor}).

If no successor wrote its TInd, then $t$ attempts to swap the field \emph{tail} back from its TInd to NONE in \lref{MCS-CAS:swapBack}. If it fails, then it has a successor, in which case $t$ waits for it to write its TInd for at most a platform-dependent period of time (\llref{MCS-CAS:waitSucStart}{MCS-CAS:waitSucEnd}) and then re-checks its \emph{next} field; if it's non-empty, $t$ signals the successor (\llref{MCS-CAS:checkSuccessor}{MCS-CAS:signalSuccessor2}).

Before exiting the CAS method, $t$ increments its \emph{modeCount} field and shifts to low-contention mode if the number of CAS operations it applied in high-contention mode reached a platform-dependent threshold value (\llref{MCS-CAS:incModeCounter}{MCS-CAS:shiftToLowContentionMode}).

When $t$ is in low-contention mode its \emph{modeCount} field is used for counting the number of consecutive CAS failures. If the current CAS operation was successful, then $t$ resets \emph{modeCount} field (\lref{MCS-CAS:resetModeCount}). If the CAS failed then the field is incremented. If the number of consecutive failures now reaches a platform-dependent threshold value, $t$ shifts to high-contention mode (\llref{MCS-CAS:shiftToContentionModeStart}{MCS-CAS:shiftToContentionModeEnd}).

\section{Appendix B: the Array-Based CAS algorithm}


\begin{algorithm}[H]
\scriptsize
\caption{\label{alg:arrayBased}The \texttt{ArrayBasedCAS} class.}

public class \textbf{AB-CAS$<$V$>$} extends AtomicReference$<$V$>$\;

\BlankLine

\Private \Class \textbf{ThreadRecord} \Proc{
	\Long \emph{modeCount}\;
	\Bool \emph{contentionMode}\;
	\Volatile \Bool \emph{request}\;
}

\BlankLine

\Private ThreadRecord[] \emph{tRecords} = \New ThreadRecord[MAX\_THREADS]\;
\Private AtomicInteger \emph{owner} = \New AtomicInteger(NONE)\;

\BlankLine

%

\Public V \textbf{read}() \Proc{
	ThreadRecord r = \emph{tRecords}[TInd]\;
	\If{r.contentionMode $\land$ (owner.get() $\neq$ \emph{TInd}) 		\nllabel{AB-CAS:ifHighContentionAndNotOwner}}
    {
    	r.\emph{request} = \True \nllabel{AB-CAS:setNotify}\;
  		\For{i=0; (i$<$MAX\_WAIT) $\land$ r.request; i++ \nllabel{AB-CAS:waitLoopStart}}
     	{
        	\If{owner.get() == NONE $\land$ owner.compareAndSet\emph{(NONE, TInd)}}
            {
				r.\emph{request} = \False \nllabel{AB-CAS:resetNotify}, \Break
            }
        }\nllabel{AB-CAS:waitLoopEnd}
        \lIf{r.request} {
        	r.\emph{request} = \False\nllabel{AB-CAS:resetNotifyIfNeeded}
        }
    }
	\Return get() \nllabel{AB-CAS:lastReadReturn}\;
}

\BlankLine

\Bool \textbf{CAS}(V old, V new) \Proc{
	\Bool ret = compareAndSet(old,new) \nllabel{AB-CAS:applyCAS}\;
	ThreadRecord r = \emph{tRecords}[TInd]\;

	\uIf {r.contentionMode \nllabel{AB-CAS:inContentionMode}}
	{
	
		\If{++r.modeCount $\geq$ \emph{NUM\_OPS} \nllabel{AB-CAS:incModeCount}}
		{
			r.\emph{modeCount} \nllabel{AB-CAS:resetModeCount}\;
	        r.\emph{contentionMode} = \False \nllabel{AB-CAS:exit}\;
	
	        \For{i = \emph{(TInd+1)}\%\emph{MAX\_THREADS}; i $\neq$ \emph{TInd}; i = $($i+1$)$\%\emph{MAX\_THREADS}\nllabel{AB-CAS:scanStart}}
			{
				\If{tRecords$[i]$.request}
				{
					\emph{owner}.set(\emph{i})\;
	                r.\emph{request} = \False\nllabel{AB-CAS:setRequestToNone}\;
	                \Return ret\;
	            }
	     	}\nllabel{AB-CAS:scanEnd}
	     	\emph{owner}.set(NONE)\nllabel{AB-CAS:setOwnerToNone}\;
	    }
	}

	\ElseIf{ret \nllabel{AB-CAS:successfulCAS}}
    {
    	r.\emph{modeCount} = 0 \nllabel{AB-CAS:resetModeCount1}\;
    }
	\ElseIf{++r.modeCount $\geq$ \emph{CONTENTION\_THRESHOLD} \nllabel{AB-CAS:CASFailed}}
    {
 		\nllabel{AB-CAS:ShiftToHighContentionMode}
        r.\emph{modeCount} = 0 \nllabel{AB-CAS:resetModeCount2},
        r.\emph{contentionMode} = \True
        \nllabel{AB-CAS:shiftToHighContentionMode2}\;
    }

	\BlankLine
	
	\Return ret \nllabel{AB-CAS:return}\;
}

\end{algorithm}

\newpage

Algorithm \ref{alg:arrayBased} presents the \texttt{ArrayBasedCAS} class, which implements a CAS contention management algorithm that we call \emph{array-based CAS}. Similarly to the MCS-CAS algorithm, with the array based CAS threads may apply their operations in either low-contention or high-contention mode.

Each class instance contains the following fields. The \emph{tRecords} array stores for each thread the following fields; \emph{contentionMode}, \emph{request} and \emph{modeCount}, which are used by the array based algorithm similarly to the way they are used by MSC-CAS. The \emph{owner} atomic integer stores the TInd of the current ``owner'' of the memory location or NONE if there is no such owner. At any point in time, the owner thread is the single high-contention mode thread that is permitted to perform read or CAS operations on the memory location encapsulated by the ArrayBasedCAS object without waiting.

We now describe the \texttt{read} operation. If thread $t$ is in low-contention mode or is the current owner (\lref{AB-CAS:ifHighContentionAndNotOwner}), then it simply delegates to the \texttt{get} method of the AtomicReference object to return the current reference value (\lref{AB-CAS:lastReadReturn}). If $t$ is in high-contention mode and is not the owner, then it initializes its \emph{request} entry to true (\lref{AB-CAS:setRequestToNone}) and executes the loop of \llref{AB-CAS:waitLoopStart}{AB-CAS:waitLoopEnd}, until it is either signaled, manages to become the owner, or performs a platform-dependent number of loop iterations. If $t$ is signaled or becomes the owner in the course of the loop then it immediately exits it, ensuring that its \emph{request} entry is reset in \lref{AB-CAS:resetNotify} or \lref{AB-CAS:resetNotifyIfNeeded}. After exiting the loop, $t$ returns the current reference value in \lref{AB-CAS:lastReadReturn}.

We now describe the \texttt{CAS} operation. First, the \texttt{compareAndSet} method on the AtomicReference superclass is called, passing to it the \emph{old} and \emph{new} operands (\lref{AB-CAS:applyCAS}). If $t$ is in high-contention mode (\lref{AB-CAS:inContentionMode}), then it is the current owner. An owner performs NUM\_OP (a platform-dependent value) number of CAS operations before releasing ownership. Thread $t$ increments its \emph{modeCount} field (\lref {AB-CAS:incModeCount}). If it has to release ownership, then it resets its \emph{modeCount} field and exits high-contention mode (\llref{AB-CAS:resetModeCount}{AB-CAS:exit}). It then scans the \emph{tRcords} array and notifies the next waiting thread (if any) that it now becomes the owner (\llref{AB-CAS:scanStart}{AB-CAS:scanEnd}). If no waiting thread is found, $t$ sets the value of the \emph{owner} field to NONE (\lref{AB-CAS:setOwnerToNone}).

If $t$ is not in high-contention mode, then it proceeds to update its statistics. If the current CAS was successful (\lref{AB-CAS:successfulCAS}), then $t$ resets its \emph{modeCount} field \lref{AB-CAS:resetModeCount1}. If the current CAS failed (\lref{AB-CAS:CASFailed}), thread $t$ increments its \emph{modeCount} field which counts the number of consecutive failures in low-contention mode. If this number now reaches a platform-dependent threshold value (\lref{AB-CAS:ShiftToHighContentionMode}), $t$ resets its \emph{modeCount} entry and shifts to high-contention mode (\llref{AB-CAS:resetModeCount2}{AB-CAS:shiftToHighContentionMode2}).

Regardless of whether $t$ is in high- or low-contention mode it always leaves the method by returning the CAS success/failure indication in \lref{AB-CAS:return}.

\end{document}